\documentclass[twocolumn,aps,showpacs,pra,superscriptaddress]{revtex4-1}
\usepackage{amsmath}
\usepackage{amssymb}
\usepackage{esint}
\usepackage{graphicx}
\usepackage{bbold}
\usepackage[normalem]{ulem}

\usepackage[colorlinks=true,allcolors=blue,bookmarks=false,pdfusetitle]{hyperref}

\makeatletter

\usepackage{color}

\newcommand{\sgn}{\mbox{sgn}}

\makeatother

\begin{document}
\title{Exact nonequilibrium dynamics of finite-temperature Tonks-Girardeau gases}
\author{Y.~Y.~Atas}
\affiliation{University of Queensland, School of Mathematics and Physics,
Brisbane, Queensland 4072, Australia}

\author{D.~M.~Gangardt}
\affiliation{School of Physics and Astronomy, University of Birmingham, Edgbaston,
Birmingham, B15 2TT, UK}

\author{I. Bouchoule}
\affiliation{Laboratoire Charles Fabry, Institut d'Optique, CNRS, Univesit\'e Paris
Sud 11, 2 Avenue Augustin Fresnel, F-91127 Palaiseau Cedex, France}

\author{K.~V.~Kheruntsyan}
\affiliation{University of Queensland, School of Mathematics and Physics,
Brisbane, Queensland 4072, Australia}

\date{\today}

\begin{abstract}
  Describing finite-temperature nonequilibrium dynamics of interacting
  many-particle systems is a notoriously challenging problem in quantum
  many-body physics. Here we provide an exact solution to this problem for a
  system of strongly interacting bosons in one dimension in the
  Tonks-Girardeau regime of infinitely strong repulsive interactions. Using
  the Fredholm determinant approach and the Bose-Fermi mapping we show how the
  problem can be reduced to a single-particle basis, wherein the
  finite-temperature effects enter the solution via an effective ``dressing''
  of the single-particle wavefunctions by the Fermi-Dirac occupation
  factors. We demonstrate the utility of our approach and its computational efficiency in two nontrivial out-of-equilibrium
  scenarios: collective breathing mode oscillations in a harmonic trap and
  collisional dynamics in the Newton's cradle setting involving real-time evolution in a periodic
  Bragg potential.
 \end{abstract}
 

  \maketitle

\section{Introduction}
  
  Out-of-equilibrium phenomena are as
  prevalent in natural and engineered systems as equilibrium ones.  Despite
  this, our understanding of nonequilibrium states of matter is far inferior
  to the understanding of equilibrium states governed by the broadly
  applicable foundational principles of statistical mechanics.  In recent
  years, ultracold quantum gases have emerged as a platform-of-choice for
  studying nonequilibrium dynamics of interacting quantum many-body systems
  \cite{Bloch-Dalibard-Zwerger,CazalillaRigolNJP2010,polkovnikov2011,Cazalilla2011,lamacraft2012,eisert2015quantum}.
  This is due to the fact that such gases represent nearly-ideal and highly
  controllable realisations of various models of many-body theory in which
  such dynamics can be accessed on observable time scales.  A particularly
  active area here concerned the study of quantum quenches and mechanisms of
  relaxation in one-dimensional (1D) Bose gases
  \cite{Kinoshita2006,hofferberth2007,trotzky2012,gring2012,fang2014quench}
  (see also
  \cite{Cazalilla2011,Caux2012,Iyer2012,*Iyer2013,Kormos2013,Nardis2014,Zill2015,*Zill2016,Piroli2016}
  and references therein), which, in the uniform limit, can be well
  approximated by the integrable Lieb-Liniger model \cite{Lieb-Liniger1963}
  with delta-function pairwise interactions between the particles.

The limit of infinitely strong repulsive interactions in the Lieb-Liniger
model corresponds to a 1D gas of impenetrable (hard-core)
bosons, 
or the Tonks-Girardeau (TG) gas.  The strong interactions required for
realizing the TG gas have been achieved in ultracold atom experiments in
highly anisotropic
traps~\cite{Kinoshita2004,paredes2004tonks,Kinoshita2006,Naagerl2009}, and its
spectacular dynamics in a quantum Newton's cradle setting were observed in
Ref.~\cite{Kinoshita2006}.
The particle impenetrability in the TG gas allows one to map the problem of
many interacting bosons to an ideal (noninteracting) gas of fermions
\cite{Girardeau1960}. Remarkably, the Bose-Fermi mapping and hence the exact
integrability of the model works not only in the uniform limit but also for
inhomogeneous systems
\cite{Girardeau_Bose_Fermi,Girardeau2000,yukalov2005fermi}, which enables
accurate tests of theory against experiments that are typically performed in
harmonic traps.
Despite this, and despite the relatively long history behind the model,
theoretical studies of TG gases have so far been limited to either zero- and
finite-temperature \emph{equilibrium} properties or \emph{zero-temperature}
dynamics~\cite{Girardeau2000,GangardtMinguzziExact,Pezer,Collura2013,QuinnHaque2014,Campbell2015,Caux2016}.
\emph{Finite-temperature dynamics}, on the other hand, has not been studied
yet, which is important for accurate comparisons with experiments that are
realized at nonzero temperatures.

In this work, we develop an exact finite-temperature dynamical theory of the
TG gas applicable to arbitrary external potentials. More specifically, we
propose a computationally efficient method for calculating the dynamics
  of 
single-particle
density matrix and the corresponding momentum distribution of the gas. The method is based
on the Fredholm determinant approach and the Bose-Fermi mapping, which allows
one to solve  the dynamical  many-body problem in terms of 
the dynamics of single-particle quantities. 
This is similar to the zero-temperature approach of
Ref.~\cite{Pezer}, except that we take into account 
finite-temperature effects. This results in  
an effective ``dressing'' 
of the single-particle wavefunctions by the
square roots of Fermi-Dirac occupation factors. Our formalism is equally
applicable to finite-temperature equilibrium calculations, in which case it
offers significant computational advantages over the previously used
approaches based on Lenard's formula
\cite{Lenard1966,vignolo2013universal,hao2016properties}. For harmonically
trapped systems, the efficiency of our approach is further unveiled by
utilising known analytic integrals and recurrence relations between Hermite
polynomials. 

\section{One-body density matrix and its evolution at finite temperature}

\subsection{Model Hamiltonian and Bose-Fermi mapping}

We consider  a 1D gas of $N$ bosons of mass $m$,
interacting via repulsive two-body delta-function potential  
and confined by a time-dependent one-body trapping potential $V(x,t)$ 
described by the Hamiltonian
\begin{equation}
\label{eq:ham}
\hat{H}=\sum_{j=1}^{N}\left[-\frac{\hbar^{2}}{2m}\frac{\partial^{2}}{\partial
    x^{2}_{j}}+V(x_j,t)\right] +g\sum_{j<l}\delta(x_{j}-x_{l}).
\end{equation}
where $g>0$ is the interaction strength. 
The  infinitely strong contact interactions
($g\rightarrow \infty$) correspond to the TG gas of impenetrable
bosons \cite{Girardeau1960,Lieb-Liniger1963}. In this limit,  the interactions 
are   replaced by the hard-core constraints  and  the  quantum many-body
problem can be solved exactly.

Our goal is to study the real-time evolution of the one-body density matrix of the TG gas,
\begin{align}
\notag \rho(x,y;t)&=\frac{1}{{\cal{Z}}}\sum_{N,\alpha} e^{\beta(
 \mu N- E_{\alpha} )}  \int dx_{2}\dots dx_{N}\\
			& \times
                        \Psi_{\alpha}(x,x_{2},\dots,x_{N};t)
                        \Psi^{\ast}_{\alpha}(y,x_{2},\dots,x_{N};t)\, .
                        \label{density-matrix-general}
\end{align}
 Here, ${\cal{Z}}=\!\sum_{N,\alpha} e^{\beta(
 \mu N- E_{\alpha} )} $ is the grand-canonical partition function, $\beta \equiv 1/k_BT_0$,
 where $T_0$ is the initial equilibrium temperature, 
 $\mu$ is the initial chemical potential, and $\Psi_{\alpha}(x_{1},...,x_{N};t)$ is the $N$-body wavefunction evolved according to the Schr\"{o}dinger equation from the initial wavefunction $\Psi_{\alpha}(x_{1},...,x_{N};0)$.

 At time $t\!=\!0$, Eq.~(\ref{density-matrix-general}) describes 
the initial thermal equilibriums state of the system  
in the trapping potential $V(x,0)$ at temperature $T_0$.
The density matrix  allows
one to calculate important observables, such as the real-space
density $\rho(x,t) \!=\! \rho(x,x;t)$ and the momentum distribution
$n(k,t)\!=\!\int dx\, dy \, e^{-ik(x-y)}\rho(x,y;t)$ of the gas.

The reduction of the many-body dynamical problem of a TG gas to a single particle
evolution relies on the existence of a Bose-Fermi
 mapping \cite{Girardeau1960,Girardeau2000,yukalov2005fermi,BuljanGasenzer2008},
\begin{equation}
\Psi_{\alpha}(x_{1},...,x_{N};t)\!=\!A(x_{1},...,x_{N})
\Psi_{\alpha}^{F}(x_{1},...,x_{N};t), 
\label{Bose_Fermi_mapping}
\end{equation} 
between the many-body wavefunctions $\Psi_{\alpha}$ of \emph{interacting}
(hard-core) bosons and those of \emph{free} fermions, $\Psi_{\alpha}^{F}$,   
where the function
$A(x_{1},...,x_{N})=\prod_{1\leq j<i \leq N}\sgn(x_{i}-x_{j})$ ensures the symmetrization of the bosonic wavefunctions.

The  fermionic wavefunctions are constructed as
Slater determinants
$\Psi_{\alpha}^{F}(x_{1},...,x_{N};t)\!=\!\mathrm{det}_{i,j=1}^{N}\left[
  \phi_{\alpha_i}(x_{j},t)\right]/{\sqrt{N!}}$
of single-particle wavefunctions $\phi_{\alpha_i} (x,t)$ evolving according to
the Schr\"{o}dinger equation, with the initial wavefunctions
$\phi_{\alpha_i} (x,0)$ being the eigenstates of the trapping potential
$V(x,0)$, with eigenenergies $E_{\alpha_i}$ such that
$E_\alpha=\sum_{i=1}^{N}E_{\alpha_i}$ and the index
$\alpha=\{\alpha_1,...,\alpha_N\}$ representing the set of single-particle
quantum numbers $\alpha_i$ that may occur.

As was shown by Lenard~\cite{Lenard1966}, the Bose-Fermi mapping 
allows one to express the one-body density matrix 
(\ref{density-matrix-general}), in terms of the fermionic one-body density matrix,
\begin{equation}
\rho_{F}(x,y;t)= {\large{\sum}}_{i=0}^{\infty}f_{i}\,\phi_{i}(x,t)\phi_{i}^{\ast}(y,t), 
\label{fermionic_kernel}
\end{equation}
which is a 
sum of products of single-particle wavefunctions weighted by the
Fermi-Dirac occupation factors $f_{i}\!=\![e^{(E_{i}-\mu)/k_BT_0}+1]^{-1}$ for the $i$th single-particle orbital ($i\!=\!0,1,...$) of energy $E_i$.
The resulting expression for $\rho(x,y;t)$ can be expressed as an infinite series 
\begin{eqnarray}
\notag \rho(x,y;t) =  {\large{\sum}}_{j=0}^{\infty}\frac{(-2)^{j}}{j!}\left[\mathrm{sign}(x-y)\right]^{j} \,\,\,\,\,\,\,\,\,\,\,\,\,\,\,\,\,\, \\
\times \int_{x}^{y}\! dx_{2}\cdots dx_{j+1}\, \mathrm{det}^{j+1}_{k,l=1}\left[ \rho_{F}(x_{k},x_{l};t)\right], 
\label{Tonks_one_body_density_matrix}
\end{eqnarray}
where 
in the determinant 
one has to take $x_{k}\!=\!x$ for $k\!=\!1$ and $x_{l}\!=\!y$ for $l\!=\!1$; the $j\!=\!0$ term in the sum is given by $\rho_F(x,y;t)$ itself. In practice, it is difficult to use this formula for increasingly higher $j$ (for example, in Ref.~\cite{vignolo2013universal} only $j\!\leq$3 terms were included in the calculated examples) as the large-$j$ terms contain multiple ($j$-fold) integrals, in addition to entering the sum with alternating signs that lead to numerical inaccuracies.

\subsection{Fredholm determinant approach to calculating the one-body density matrix}

Here, we instead follow the approach of Refs.~\cite{forrester2003painleve,jimbo1980density}, which identified an alternative and more compact form of Lenard's formula, given by  
\begin{equation}
\rho(x,y;t)=\mathrm{det}\left( 1-2\hat{K}(t)\right)R(x,y;t), \label{Fredholm_OBDM}
\end{equation}
i.e., a product of a Fredholm determinant and the associated resolvent operator $R(x,y;t)$ 
of the integral operator $\hat{K}$, whose action  
on an arbitrary function $g(r)$ is given 
by $(\hat{K}g)(w)\!=\!\int_{x}^{y}K(w,r;t)g(r)dr$, with the kernel $K(w,r;t)\!=\! \rho_F(w,r;t)$ in our case.
The resolvent operator $R(x,y;t)$ satisfies the following integral equation \cite{jimbo1980density}: 
\begin{equation}
R(u,v;t)-2\int_{x}^{y}K(u,r;t)R(r,v;t)dr=K(u,v;t). 
\label{Resolvent_integral_equation}
\end{equation}
Here, we have assumed $y\geq x$ without loss of generality and suppressed, for notational simplicity, the dependence of $R$ on the integration limits as the final results that we are interested in only depend on the values of $R$ at $u\!=\!x$ and $v\!=\!y$.
 We point out that Eq.~(\ref{Tonks_one_body_density_matrix}) corresponds to the expansion of the determinant in Eq.~(\ref{Fredholm_OBDM}) 
  by minors \citep{bornemann2010numerical,forrester2003painleve}, and that a discrete version of Eq.~(\ref{Fredholm_OBDM}) 
  on a lattice has previously been obtained by Y. Castin for a spatially homogeneous TG gas at $T=0$ (see Eq.~(3.37) in \cite{Castin2004simple}).

At zero temperatures, the infinite sum appearing in the fermionic one-body density matrix (\ref{fermionic_kernel}), which also serves the role of the kernel $K$ in Eq.~(\ref{Resolvent_integral_equation}),
  is effectively truncated by the highest occupied orbital term ($i\!=\!N\!-\!1$) corresponding to the Fermi level. At finite temperatures this is no longer true; however, for any practical calculation the infinite series can be truncated at some large $M$ beyond which the Fermi-Dirac occupancies are negligible. (In practice, the precise value of the cutoff $M$ should be determined from the convergence properties of the final physical results of interest.)
Therefore, to a good approximation, the fermionic kernel in Eq.~(\ref{Resolvent_integral_equation}) 
can be replaced by a finite series $\rho_{F}(w,r;t)\simeq K_{M}(w,r;t)=\sum_{i=0}^{M} f_{i}\phi_{i}(w,t)\phi_{i}^{\ast}(r,t).$ Inserting this form of the kernel into Eq.~(\ref{Resolvent_integral_equation}) 
gives
\begin{equation}
R(u,v;t)=K_{M}(u,v;t)+2{\large{\sum}}_{i=0}^{M}\sqrt{f_{i}}\phi_{i}(u,t)A_{i}(v;t), 
\label{Resolvent_sum}
\end{equation}
where we have introduced the following notation,
\begin{equation}
A_{i}(v;t)=\sqrt{f_{i}}\int_{x}^{y}\!\phi_{i}^{\ast}(r,t)R(r,v;t)\,dr. 
\end{equation}

The functions $A_{i}(v;t)$ are determined as follows. Multiplying Eq.~(\ref{Resolvent_sum}) by $\sqrt{f_{j}}\phi_{j}^{\ast}(u,t)$ and integrating on $[x,y]$, we obtain
\begin{equation}
A_{j}(v;t)={\large{\sum}}_{i=0}^{M}S_{ji}(t)\left[\sqrt{f_{i}}\phi_{i}^{\ast}(v,t)+2 A_{i}(v;t)\right], \label{Equation_of_A}
\end{equation}
where the matrix elements $S_{ij}=\left(\mathbf{S} \right)_{ij}$ are given by
\begin{equation}
S_{ij}(t)=\mathrm{sign}(y-x)\sqrt{f_{i}f_{j}}\int_{x}^{y}\!\phi_{i}^{\ast}(x^{\prime}\!,t)\phi_{j}(x^{\prime}\!,t)\,dx^{\prime}, 
\end{equation}
and where we again suppressed the dependence of $S_{ij}(t)$ on the integration limits.

We proceed by writing the equation satisfied by the functions $A_{i}$ in a more compact matrix form. By writing the left-hand side of Eq.~(\ref{Equation_of_A}) as $A_{j}(v;t)=\sum_{i=0}^{M}\delta_{ji}A_{i}(v;t)$, we obtain
\begin{equation}
{\large{\sum}}_{i=0}^{M}\!\left[ \delta_{ji}-2 S_{ji}(t)\right]A_{i}(v;t)\!=\!{\large{\sum}}_{i=0}^{M}\!S_{ji}(t)\sqrt{f_{i}}\phi_{i}^{\ast}(v,t). 
\end{equation}
Introducing the vectors $\vec{A}\!=\!\left(A_{0},...,A_{M}\right)^{\mathsf{T}}$ and $\vec{\Phi}\!=\!\left(\sqrt{f_{0}}\phi_{0}^{\ast},...,\sqrt{f_{M}}\phi_{M}^{\ast}\right)^{\mathsf{T}}$, this can be rewritten as a matrix equation, $\left[ \mathbb{1}-2 \mathbf{S}(t) \right] \vec{A}(v;t)\!=\!\mathbf{S}(t) \vec{\Phi}(v,t)$, which in turn can be inverted to yield $\vec{A}(v;t)\!=\!\left[ \mathbb{1}-2 \mathbf{S}(t) \right]^{-1} \mathbf{S}(t) \vec{\Phi}(v,t).$
Inserting this expression into Eq.~(\ref{Resolvent_sum}) and rewriting the fermionic kernel as a double sum, $K_{M}(u,v;t)=\sum_{i,j} [\sqrt{f_{i}} \phi_{i}(u,t) \,\delta_{ij} \sqrt{f_{j}} \phi_{j}^{\ast}(v,t)]$, we obtain that the resolvent operator $R(x,y;t)$ is given by
\begin{equation}
R(x,y;t)\!=\!\!{\large{\sum}}_{i,j=0}^{M}\!\!\sqrt{f_{i}}\,\phi_{i}(x,t)\left(\mathbb{1}-2 \mathbf{S}^{-1}\right)_{ij}\!\!\sqrt{f_{j}}\,\phi_{j}^{\ast}(y,t). \end{equation}

The Fredholm determinant that appears in the definition of the one-body density matrix, given by Eq.~(\ref{Fredholm_OBDM}),
 is equal to $\mathrm{det}\left(\mathbb{1}-2 \mathbf{S} \right)$ in the truncated basis \cite{bornemann2010numerical}.
Therefore, the corresponding final expression for the one-body density matrix of a finite-temperature TG gas, after taking the limit $M\!\rightarrow \!\infty$, can be written as
\begin{equation}
\rho(x,y;t)={\large{\sum}}_{i,j=0}^{\infty}\sqrt{f_{i}}
\phi_{i}(x,t)Q_{ij}(x,y;t)\sqrt{f_{j}}\phi_{j}^{\ast}(y,t), 
\label{Finite_temp_densitymat_Tonks}
\end{equation}
Here, $Q_{ij}$ are the  
matrix elements of the operator 
$\mathbf{Q}(x,y;t)=(\mathbf{P}^{-1})^{\mathsf{T}}\mathrm{det}\;\!\mathbf{P}$ (which is an $M\!\times \!M$ matrix in the truncated basis), with
\begin{equation}
 P_{ij}(x,y;t)\! =\! \delta_{ij}- 2\;\!\sgn(y-x)\sqrt{f_{i}f_{j}} \!\int_{x}^{y}
 \!\!\!dx^{\prime}\phi_{i}(x^{\prime}\!,t)\phi_{j}^{\ast}(x^\prime\!,t).
  \label{matrix_element}
\end{equation}

Thus, we have reduced Eq.~(\ref{Fredholm_OBDM}) to a
 simple double sum, which does not contain multiple integrals or sign-alternating terms present in Lenard's formula.
At zero temperature, Eqs.~(\ref{Finite_temp_densitymat_Tonks}) and (\ref{matrix_element}) reduce to the results of Ref. \cite{Pezer} as the Fermi-Dirac distribution function in this case is given by a step function equal to $1$ for orbitals with $i \!\leqslant \!N-1$, 
or $0$ otherwise. 
At nonzero temperature, the orbital wavefunctions, as our results show, become ``dressed'' by the square roots of the Fermi-Dirac occupation factors, ensuring, e.g., that the correct real-space density $\rho(x,t)\!\equiv \!\rho(x,x;t)\!=\!\sum_{i=0}^{\infty} \! f_{i}|\phi_{i}(x,t)|^{2}$ is recovered.

Equations (\ref{Finite_temp_densitymat_Tonks}) and (\ref{matrix_element}) are the main results of this paper, representing a compact and computationally practical recipe for calculating the time-dependent one-body density matrix of the TG gas. They reduce the problem of finding $\rho(x,y;t)$ to solving the time-dependent Schr\"{o}dinger
equation for the single-particle orbitals $\phi_j(x,t)$ and calculating the matrix elements
$P_{ij}(x,y;t)$.  At time $t\!=\!0$, Eq.~(\ref{Finite_temp_densitymat_Tonks})
describes the initial finite-temperature equilibrium one-body density matrix; in its present form
it offers a more efficient and accurate way of calculating $\rho(x,y;0)$
compared to the previous approaches \cite{Lenard1966,vignolo2013universal}.

\subsection{Dynamics in a harmonic trap}

The calculation of the one-body density
matrix $\rho(x,y;t)$, given by Eq.~(\ref{Finite_temp_densitymat_Tonks}), requires, in
general, the evaluation of the overlap matrix elements $P_{ij}(x,y;t)$, given by 
Eq.~(\ref{matrix_element}), between the time-evolved wave functions
$\phi_j(x,t)$, starting from the initial single-particle wave functions
$\phi_j(x,0)$.  For the special case of evolution in a time-dependent
\emph{harmonic} trap, $V(x,t) = m\omega(t)^2 x^2/2$, the wavefunctions
$\phi_j(x,0)$ are given by the well-known Hermite-Gauss orbitals, whereas the
evolution under the single-particle Schr\"{o}dinger equation can be solved using a
scaling transformation \cite{PerelomovBook,GangardtMinguzziExact}, which in
turn leads to
\begin{equation}
\rho(x,y;t)=\frac{1}{\lambda}
\rho_0\left(x/\lambda,y/\lambda\right) e^{i m\dot\lambda (x^2-y^2)/2\hbar \lambda}
\label{eq.ss},
\end{equation}
where $\rho_0(x,y)=\rho(x,y;0)$ is the initial one-body density matrix.  The scaling
parameter $\lambda(t)$ is determined from the solution of the second-order
ordinary differential equation (ODE), $\ddot{\lambda}=-\omega(t)^{2}\lambda+\omega_{0}^{2}/\lambda^{3}$, 
with the initial conditions $\lambda(0)\!=\!1$, and $\dot{\lambda}(0)\!=\!0$.  
For the quench of the trapping frequency considered above,
this ODE acquires the form of the Ermakov-Pinney equation,
$\ddot{\lambda}=-\omega_1^{2}\lambda+\omega_{0}^{2}/\lambda^{3}$, with the
solution $\lambda(t) \!=\! [1 +\epsilon \sin^2(\omega_1 t)]^{1/2}$,

The scaling solution (\ref{eq.ss}) enormously simplifies  the calculation of $\rho(x,y;t)$ as Eq.~(\ref{Finite_temp_densitymat_Tonks}) is used 
only once---for calculating the \emph{initial} equilibrium density matrix $\rho_0(x,y)$ of a harmonically trapped TG gas.
In this case, the elements of the overlap matrix $P_{ij}(x,y;0)$ are computed for the harmonic oscillator eigenstates, $\phi_{j}(x)= e^{-x^2/2l_{\mathrm{ho}}^2}H_{j}(x/l_{\mathrm{ho}})/(\pi^{1/4}\sqrt{2^jj!l_{\mathrm{ho}}})$,
where $H_j(\xi)$ is the Hermite polynomial of degree $j$ ($j=0,1,2,...$), and $l_{\mathrm{ho}}=\sqrt{\hbar/m\omega_0}$ is the harmonic oscillator length. One then computes the determinant of the initial overlap matrix $\mathbf{P}$ and  inverts it in order to evaluate the matrix elements $Q_{ij}(x,y,0)$ appearing in Eq.~(\ref{Finite_temp_densitymat_Tonks}).

In order to describe higher-temperature samples and larger total number of atoms $N$ with this seemingly straightforward procedure, one needs to incorporate increasingly higher orbital wave functions in the double sum in Eq.~(\ref{Finite_temp_densitymat_Tonks}). This, in turn, requires evaluation of the overlap integrals between highly excited states in Eq.~(\ref{matrix_element}). (For example, for our highest temperature and highest $N$ samples, we used harmonic-oscillator excited states of up to $j\!=\!400$.) As the highly excited states are fast oscillating functions in position space, brute-force numerical integration will result in computational difficulties.

To overcome these difficulties, we instead develop and compute the overlap matrix elements using an alternative approach. Namely, for the off-diagonal elements, $P_{jk}(x,y;0)$ ($j\neq k$), we resort to a known analytic formula for the harmonic-oscillator eigenstates, given in the form of the following indefinite integral \cite{Piquette}:
\begin{equation}
\int \!\!\varphi_{j}(\xi) \varphi_{k}^{\ast}(\xi)d\xi =\frac{\mathrm{e}^{-\xi^{2}} [H_{j+1}(\xi)H_{k}(\xi) \!-\!H_{j}(\xi)H_{k+1}(\xi)]}{2(k-j)\sqrt{2^{j+k}\pi\, j!\,k!}}, 
 \label{eq:off-diagonal}
\end{equation}
where $\xi\!\equiv \!x/l_{\mathrm{ho}}$ and $\varphi_j(\xi)\!\equiv \!\sqrt{l_{\mathrm{ho}}} \phi_j(x)$.
This formula is much simpler to use, especially at higher temperatures and larger $N$, than the one based on a finite series of confluent hypergeometric functions used in Ref.~\cite{vignolo2013universal}.

For the diagonal elements $P_{jj}(x,y;0)$, no similar  formula exists to the best of our knowledge, however, we find that these elements can be computed efficiently using the following recursive method. We define a sequence of functions $\lbrace M_{j}(\xi) \rbrace_{j=0,1,\dots}$ containing the desired diagonal matrix elements in the form of indefinite integrals,
\begin{equation}
M_{j}(\xi)=\frac{\sqrt{\pi}}{2}\mathrm{erf}(\xi)-\frac{1}{2^j j!}\int \mathrm{e}^{-\xi^2}H^{2}_{j}(\xi)\mathrm{d}\xi, 
\label{eq:M}
\end{equation}
where $\mathrm{erf}(\xi)$ is the error function and $M_{0}(\xi)=0$. Using the well-known recurrence relation for the Hermite polynomials, this yields\begin{equation}
M_{j+1}(\xi)=M_{j}(\xi)+\frac{\mathrm{e}^{-\xi^2}}{2^{j+1}(j+1)!}H_{j}(\xi)H_{j+1}(\xi). 
\label{eq:M-reccurence}
\end{equation}
Equations (\ref{eq:off-diagonal})--(\ref{eq:M-reccurence}) thus allow for an efficient computation of all (diagonal and off-diagonal) matrix elements of $P_{ij}(x,y;0)$ without performing explicit numerical integration of products of harmonic oscillator wavefunctions.

\section{Examples of evolution of the Tonks-Girardeau gas from a thermal equilibrium state}

As an immediate application and illustration of the broad applicability of our  approach, we use it to analyze
two paradigmatic problems of current experimental and theoretical interest: (a) collective breathing-mode oscillations of a finite-temperature TG gas in a
harmonic trap, and (b) collisional dynamics in the Newton's cradle setting  which
involves real-time evolution in a periodic Bragg potential.

\begin{figure}[tbp]
\includegraphics[width=8.6cm]{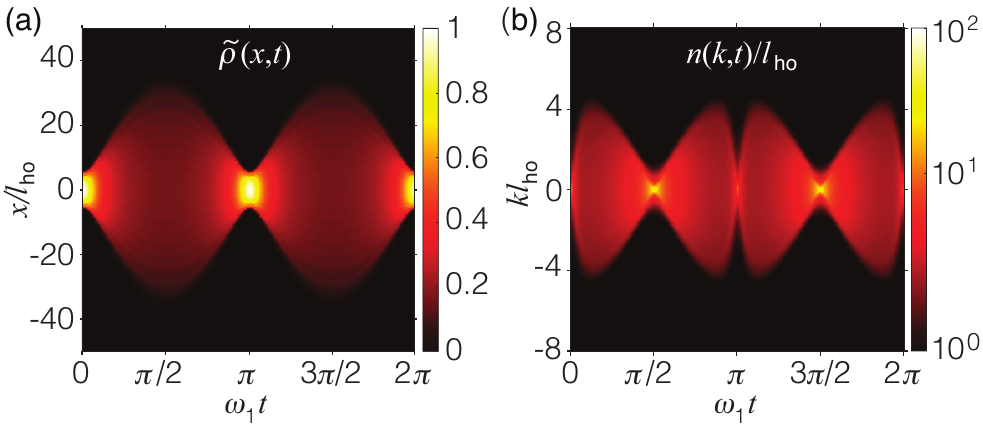}
 \caption{(Color online) Breathing-mode dynamics of the TG gas following a confinement 
     quench.  (a) Real-space density $\tilde{\rho}(x,t)\equiv\rho(x,t)/\rho(0,0)$ and (b) momentum distribution, $n(k,t)/l_{\text{ho}}$ (where $l_{\mathrm{ho}}\!=\!\sqrt{\hbar/m\omega_0}$ is the harmonic oscillator length) as functions of the dimensionless time $\omega_1t$, for $N\!=\!16$ particles, quench strength $\epsilon\!=\!35$, and dimensionless initial temperature $\theta_0 \!\equiv \!k_{B}T_0/N\hbar\omega_{0}\!=\!0.01$.} 
     \label{fig:breathing}
\end{figure}

For the
first application, we consider a TG gas initially in thermal equilibrium in a harmonic potential $V(x,0)\!=\!
m\omega_0^2 x^2/2$ with the frequency $\omega_0$. To invoke the 
breathing-mode oscillations we use a confinement quench in which at $t\!=\!0$ 
the trap frequency is instantaneously changed  
from the pre-quench value $\omega_0$ to a new value $\omega_1$; we characterise 
the quench strength by a dimensionless parameter $\epsilon\!=\!\omega_0^2/\omega_1^2-1$. 
Figure~\ref{fig:breathing} shows 
the evolution of the density profile $\rho(x,t)$ and the momentum
distribution $n(k,t)$ after a strong quench
($\epsilon\!=\!35$),  for $N\!=\!16$ particles and a dimensionless
initial temperature of $\theta_0\!\equiv \!k_BT_0/N\hbar \omega_0\!=\!0.01$.  As follows from the scaling solutions of Eq.~(\ref{eq.ss}),
the dynamics of $\rho(x,t)$  consists of self-similar
broadening and narrowing cycles  
occurring at the fundamental breathing-mode frequency of
$\omega_B\!=\!2\omega_1$.
In contrast, the momentum
distribution displays periodic broadening and narrowing cycles that occur at \emph{twice}
the rate of the oscillations of the \emph{in situ} density profile. Unlike the breathing-mode
oscillations of an ideal Fermi gas, the momentum distribution of the TG gas
becomes narrow not only at the outer turning points of the classical harmonic
oscillator motion, when the \emph{in situ} density profile is the broadest
(here corresponding to time instances of $\omega_1t\!=\!\pi/2+\pi l$, with
$l\!=\!1,2,...$), but also at $\omega_1t\!=\!\pi l$ when the gas is maximally
compressed. We refer to these points as the \emph{inner} turning points, which
serve as a manifestation of a collective many-body bounce effect due to the
increased thermodynamic pressure of the gas that acts as a potential
barrier. This phenomenon is similar to frequency doubling observed recently in
a weakly-interacting quasicondensate
regime~\cite{fang2014quench,Bouchoule_qbec:2016} and is further explored
in Ref.~\cite{TonksBounce}.

\begin{figure}[tbp]
  \includegraphics[width=7.2cm]{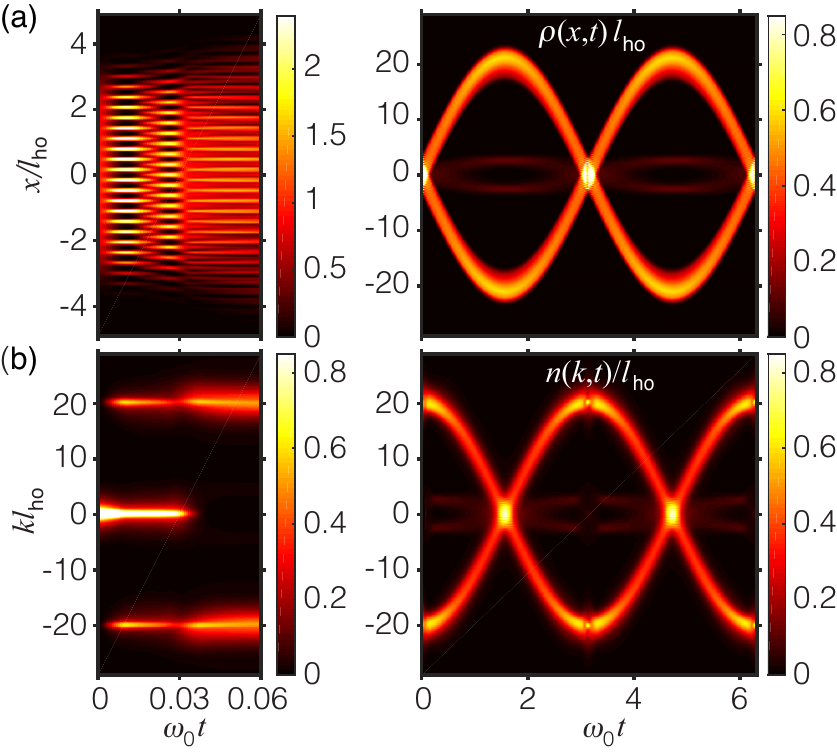} \caption{ (Color
    online) Dynamics of the TG gas in the Newton's cradle setting. (a) The evolution of the real-space density,
    $\rho(x,t)l_{\mathrm{ho}}$, as a function of the dimensionless time
    $\tau=\omega_0t$; the left panel is the magnified view into the time window
    containing the Bragg pulse sequence \cite{Bragg-pulse}, whereas the right panel
    shows the full time window including post-Bragg periodic oscillations in
    the purely harmonic potential. (b) The respective momentum
    distribution, $n(k,t)/l_{\text{ho}}$. In this example, $\theta_0\!=\!0.1$,
    $N\!=\!5$, and $k_0l_{\mathrm{ho}}\!=\!10$. }
\label{fig:NC}
\end{figure}

As a second application of our approach, we analyze the dynamics of a
finite-temperature TG gas in the Newton's cradle setting
\cite{Kinoshita2006}. In this example (see Fig.~\ref{fig:NC}), the initial atomic cloud in thermal
equilibrium at temperature $\theta_0\!=\!0.1$ is subjected to a sequence of
laser induced Bragg pulses optimized to split the atomic wavepacket into two
counter-propagating halves corresponding to $\pm 2\hbar k_0$ diffraction
orders of Bragg scattering~\cite{Wu_Bragg_pulse:2005}. This is modelled by a
periodic lattice potential $V_{\mathrm{B}}(x,t)\!=\!\Omega(t)\cos(2k_0x)$ of an
amplitude $\Omega(t)$ (consisting of two square pulses \cite{Bragg-pulse}), superimposed on top of the initial harmonic 
potential of frequency $\omega_0$.
Unlike the (short pulse) Kapitza-Dirac regime of Bragg scattering analyzed, e.g., in
Ref.~\cite{Caux2016}, we operate in the (long pulse) Bragg regime of the
Newton's cradle experiment~\cite{Kinoshita2006} wherein the interatomic
interactions during the Bragg pulse are automatically taken into account,
rather than neglected. The subsequent collisional dynamics of the gas
in the underlying pure harmonic trap potential displays periodic behavior and the characteristic traits 
observed in~\cite{Kinoshita2006}.

\begin{figure}[t]
\includegraphics[width=8.5cm]{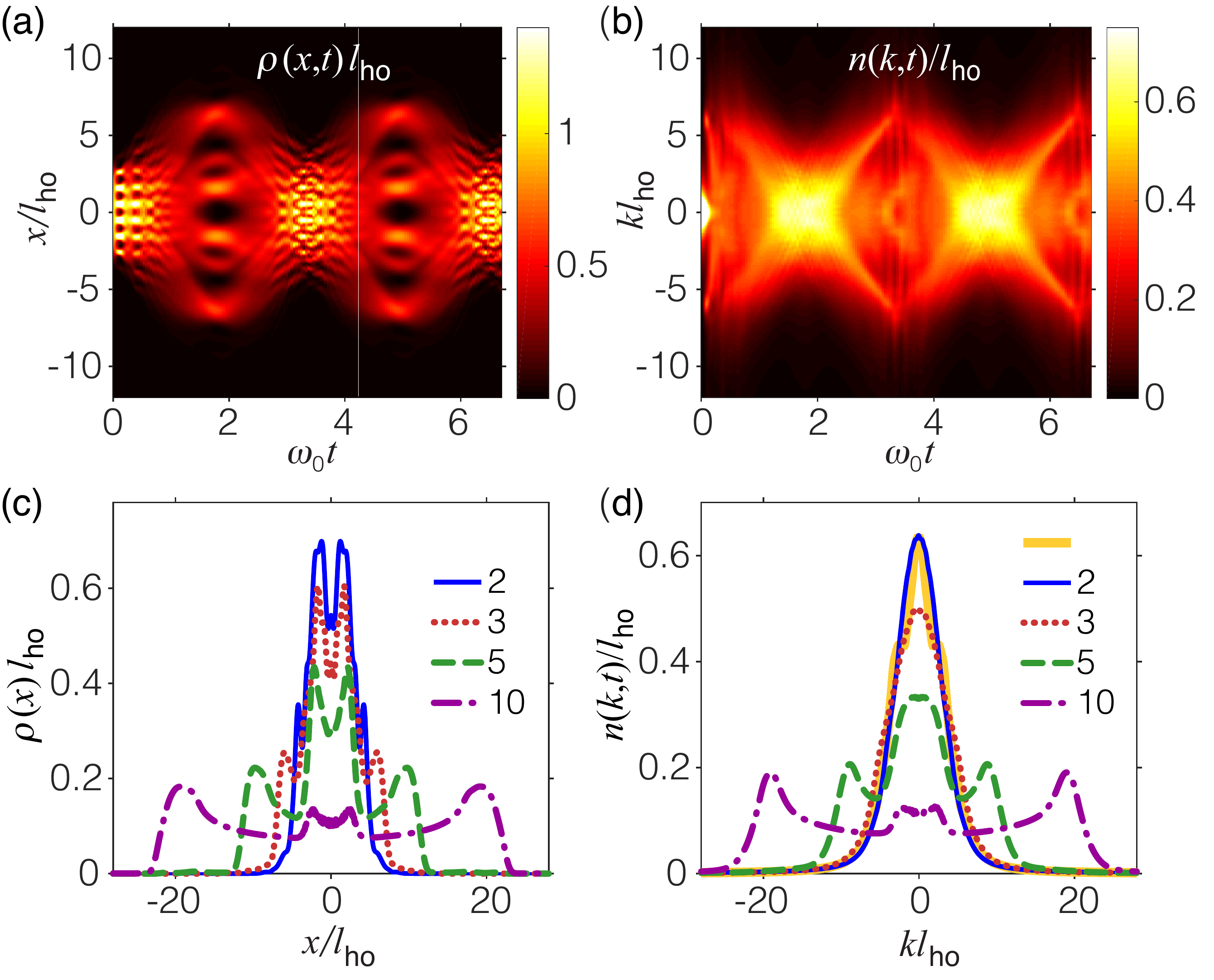}
 \caption{(Color
    online) (a), (b) Same as in the main panels of Fig.~\ref{fig:NC}, but for $k_0l_{\mathrm{ho}}\!=\!3$. (c), (d) The real-space density and momentum distributions averaged over a full oscillation period (as in Ref.~\cite{Kinoshita2006}) starting immediately after the end of the Bragg pulse at time $t_B$, for $k_0l_{\mathrm{ho}}\!=\!2,3,5,10$. In (d), the thick (light orange) solid line shows the momentum distribution $n(k,t_B)$, for $k_0l_{\mathrm{ho}}\!=\!2$.
 }
\label{fig:NC-extra}
\end{figure}

In Fig.~\ref{fig:NC-extra}, we show the collisional dynamics 
under the same initial conditions, but for a smaller Bragg momentum. This is essentially equivalent to considering
a higher temperature sample and the same Bragg momentum as before: when
the Bragg momentum becomes comparable to the initial thermal width of the momentum distribution, 
the Bragg pulse no longer splits the distribution into two well-defined peaks. As a result, we observe a rather distorted pattern of collisional oscillations, which nevertheless display the same periodicity as previously.

\section{Summary}

In conclusion, we have developed an exact finite-temperature dynamical theory of the Tonks-Girardeau gas applicable to arbitrary initial temperatures and trapping potentials, including arbitrary variations of the trapping potentials with time.  The approach relies on the Fredholm determinant representation and the Bose-Fermi mapping, allowing one to reduce the problem of many-body evolution to a single-particle basis. For harmonically trapped gases, the approach further benefits from analytic scaling solutions for the single-particle wave functions, while for arbitrary trapping potentials the wave functions should be evolved numerically according to the single-particle Schr\"{o}dinger equation. 
Our results open the way to systematic studies of nonequilibrium dynamics of this paradigmatic strongly interacting many-body system. The examples illustrated here concerned the breathing-mode oscillations and the Newton's cradle setup; however, other nonequilibrium scenarios can be easily considered,  such as periodic driving, collisions in anharmonic traps, and formation of quantum shock waves, to name a few. In addition, our approach can be extended  to treat finite-temperature dynamics of related integrable models, such as the $XY$ spin model \cite{lieb1961two}.
  
\begin{acknowledgments}
The authors acknowledge fruitful discussions with Y. Castin, E. Bogomolny, and O. Giraud. Y.\,Y.\,A. 
thanks R.\,J.\, Lewis-Swan for the introduction to the XMDS software package used in the numerical simulations of the single-particle Schr\"{o}dinger equation.
I.\,B. acknowledges support by the Centre de Comp\'{e}tences Nanosciences \^{I}le-de-France.
K.\,V.\,K. acknowledges support by the Australian Research Council Discovery Project Grant,
 Grant No. DP140101763.
\end{acknowledgments}


%

\end{document}